\documentclass[12pt]{iopart}

\newcommand{\ca}{Ca$^{2+}$}  
\newcommand{\ip}{IP$_3$}
\newcommand{\ipr}{IP$_3$R}
\newcommand{\iprs}{IP$_3$Rs}

\usepackage{iopams}  
\usepackage{color}
\usepackage{graphicx} % Required for inserting images

\begin{document}

\title[Frequency encoding and excitability]{Cell information processing via frequency encoding and excitability}

\author{Alan Givré \& Silvina Ponce Dawson}

\address{UBA, FCEN, Departamento de F\'\i sica \&  CONICET-UBA, IFIBA,\\ Ciudad Universitaria Pab. I, (1428) Buenos Aires, Argentina}
\ead{silvina@df.uba.ar}
\vspace{10pt}
\begin{indented}
\item[]April 2024
\end{indented}

\begin{abstract}
{Cells continuously interact with their environment mediating their
responses through {\it signaling cascades}. Very often, external
stimuli induce pulsatile behaviors in intermediaries of the cascade of
increasing frequency with the stimulus strength. This is
characteristic of intracellular \ca\ signals involving \ca\ release
through Inositol Trisphosphate Receptors (\iprs). The mean frequency
of \ipr-mediated \ca\ pulses has been observed to scale exponentially
with the stimulus strength in many cell types.  In this paper we use a
simple ODE model of the intracellular \ca\ dynamics for parameters for
which there is one excitable fixed point. Including fluctuations
through an additive noise term, we derive the mean escape rate from
the stationary state and, thus, the mean interpulse time, as a
function of the fraction, $\beta$, of readily openable \iprs. Using an
\ipr\ kinetic model, experimental observations of spatially resolved
\ca\ signals and previous estimates of the \ip\ produced upon 
stimulation we quantify the fluctuations and relate $\beta$ to [\ip]
and the stimulus strength. In this way we determine that the mean
interpulse time can be approximated by an exponential function of
the latter for ranges such that
the covered mean time intervals are similar or larger than those
observed experimentally. The study thus provides an easily
interpretable explanation, applicable to other pulsatile signaling
intermediaries, of the observed exponential dependence between frequency and stimulus, a key feature 
that makes frequency encoding qualitatively different from other ways commonly used by cells to ``read'' their environment.}
\end{abstract}

%
% Uncomment for keywords
%\vspace{2pc}
%\noindent{\it Keywords}: XXXXXX, YYYYYYYY, ZZZZZZZZZ
%
% Uncomment for Submitted to journal title message
%\submitto{\JPA}
%
% Uncomment if a separate title page is required
%\maketitle
% 
% For two-column output uncomment the next line and choose [10pt] rather than [12pt] in the \documentclass declaration
%\ioptwocol
%

\section{Introduction}

Cells continuously interact with their surroundings, not only to exchange nutrients and energy, but also to detect environmental changes and act accordingly~\cite{bialek_review2016}. Very frequently changes in the environment are endowed in concentration changes of substances that can bind to plasma membrane receptors. This, in turn, induces changes in the concentration and/or activation level of various intracellular
components in what is known as a {\it signaling cascade}. Cells use different strategies to go from the external stimulus to the end response~\cite{tostevin_2012, micali_pcbi_2015,givre_sci_rep_2023}. Sometimes, increasing intensities of the external stimulus are ``encoded'' in increasing concentrations of the internal {\it messengers} (amplitude codification)~\cite{amplitude5}. On other occasions,  stimuli induce a pulsatile behavior in some of the intermediaries in which the frequency increases with  the stimulus strength (frequency encoding)~\cite{Cai2008, Carbo15022017}. This last behavior has been observed in intracellular \ca\ signals~\cite{Dupont2011,Thurley2014} that involve \ca\ release from the endoplasmic reticulum (ER) through Inositol Trisphosphate Receptors (\iprs)~\cite{Foskett2007}. 

\ca\ is ubiquitous as intracellular messenger. It is involved in the
control of many cellular functions including both fertilization and
cell death~\cite{Bootman2012}. On many occasions, the selective
control of specific targets depends on the spatio-temporal of the
cytosolic \ca\ concentration.  \ca\ release from the ER through
\iprs\ give cells the tools to produce different spatio-temporal
patterns. On one hand, \iprs\ need to bind \ip\ and \ca\ to become
open. Therefore, the opening of one such channel can lead to the
opening of many others due to {\it Calcium Induced Calcium
  Release}~\cite{Fabiato83}.  \iprs, in turn, are usually organized in
clusters containing $\sim 10-20$ channels. Furthermore, high
\ca\ concentrations inhibit \iprs.  Thus, the signals can range from
                    {\it blips} and {\it puffs} involving \ca\ release
                    from one channel or cluster, to waves that
                    propagate throughout the cell depending on the
                    level of \ip~\cite{Yao1995,sun1998,Smith2009} and
                    the interplay between \ca\ activation and
                    inhibition. {Under physiological conditions, \ip\ is made via the hydrolysis of precursors that are on
the plasma membrane by the enzyme Phospholipase C (PLC) which is
activated as a result of the binding of external effectors to specific
plasma membrane receptors.} In many different cell types, constant
                    stimulation with extracellular ligands can yield
                    sequences of these waves that, when the interior
                    of the cell is not resolved, are visualized with
                    fluorescence microscopy techniques as sequences of
                    \ca\ spikes~\cite{Dupont2011,Thurley2014}.
                    Usually each \ca\ spike is preceded by localized
                    signals which start to occur more frequently until
                    they yield the global release
                    event~\cite{Marchant2001}. This is typical of
                    spatially extended excitable systems in which
                    random local (\ca\ release) events are amplified
                    through space eventually triggering an ``extreme''
                    one (a \ca\ spike)~\cite{Lopez2012,PhysRevResearch.3.023133}.  We
                    hereby recall the defining characteristics of
                    excitable systems: the existence of a stable
                    stationary state, a threshold which, if surpassed
                    by perturbations, a long excursion in phase space
                    occurs before the system returns to the stationary
                    solution and a refractory period before a new
                    spike can be elicited by
                    perturbations~\cite{izhikevich_book}.

A systematic study of intracellular \ca\ pulses evoked by constant
concentrations of extracellular ligands showed, in many cell types,
that the interspike time interval could be approximated by the sum of
a fixed term, $T_{min}$, accounting for spike duration and
refractoriness, and a stochastic term, $T_{stoch}$, of average,
$T_{av}\equiv \langle T_{stoch}\rangle$ which depended exponentially
on the extracellular concentration~\cite{Thurley2014}. This dependence
which we have already used to study information transmission via
frequency encoding~\cite{givre_2018,Givre2023_2}, ressembles that of
thermally activated barrier crossing~\cite{Kramers}, where the
crossing rate is an exponential function of the activation energy
threshold. A similar behavior occurs for the average escape time in
noise-driven excitable systems~\cite{Lindner2004} in which the
external perturbation either increases the noise level or reduces the
excitability threshold~\cite{SISR,SISRvsCR,Eguiamindlin2000}.  The
dynamics of intracellular \ca\ has been associated with
excitability~\cite{lechleiter_science,lechleiter_cell_1992}, various
models of this dynamics are excitable~\cite{lirinzel,
  Tang7869,RAMLOW2023I,RAMLOW2023II} and have even been used to study
information transmission via frequency encoding {although under the assumption of a linear relation between frequency and stimulus strength}~\cite{Tang7869}. In
this paper we study whether the experimentally observed exponential
dependence of the average time interval between subsequent \ca\ spikes
can be derived from an excitable model of intracellular
\ca\ dynamics. Using the simple model developed by Dupont and
Goldbeter~\cite{Dupont1993} we derive the mean escape rate from the
stationary state. {Combining this result with the DeYoung-Keizer kinetic model of \iprs~\cite{dy_keizer}
and experimental observations of \ca\ signals elicited in {\it X. Laevis} oocytes~\cite{Marchant2001}}, we
 determine how the escape rate depends on the \ip\ concentration
and, consequently, on the external stimulus strength under the
assumption that it is proportional to [\ip]. This derivation provides
a theoretical framework to support our recent study which shows that
frequency and amplitude encodings equip cells with qualitatively
different tools to sense their environment~\cite{Givre2023_2}.

\section{The model}

In this work we use the very simple one-pool model of intracellular \ca\ dynamics introduced by Dupont and Goldbeter~\cite{Dupont1993}. The equations defining this model are:
\begin{eqnarray}
\dot{C}&=v_0+v_1\beta-v_{M2}\frac{C^n}{K_2^n+C^n}+\beta v_{M3}\frac{C^p}{K_A^p+C^p}\frac{Y^m}{K_R^m+Y^m}\nonumber\\&\quad+k_fY-kC,\label{eq:C_1}\\
\dot{Y}&=v_{M2}\frac{C^n}{K_2^n+C^n}-\beta v_{M3}\frac{C^p}{K_A^p+C^p}\frac{Y^m}{K_R^m+Y^m}-k_fY,\label{eq:Y}
\end{eqnarray}
where $C$ is  the free \ca\ concentration in the cytosol
and $Y$ can be assumed to be proportional to the free [\ca] in the lumen of the ER with the constant of proportionality taking into consideration the ratio of the luminal to the cytosolic volumes and of the \ca\ buffering capacities in both pools~\cite{Lopez_2016}; $v_0$ represents the \ca\ influx from the
extracellular medium; the two terms proportional to $\beta$ represent
the \ip-dependent release of \ca\ from the ER with
$0\le \beta\le 1$, the fraction of \ip-bound \iprs; the
term proportional to $v_{M2}$ represents \ca\ pumping back into the ER, the term $k_f Y$
a \ca\ leak from the ER
and the last term in Eq.~(\ref{eq:C_1}) is the removal of free cytosolic
\ca\ by various means.   To facilitate the computations, as done in~\cite{SchusterStefanMarhl}, we introduce the variable $Z\equiv C+Y$ and rewrite Eqs.~(\ref{eq:C_1})--(\ref{eq:Y}) as:
\begin{eqnarray}
\dot{C}&=v_0+v_1\beta-v_{M2}\frac{C^n}{K_2^n+C^n}+\beta v_{M3}\frac{C^p}{K_A^p+C^p}\frac{(Z-C)^m}{K_R^m+(Z-C)^m}\nonumber\\&\quad+k_f(Z-C)-kC,\label{eq:C}\\
\dot{Z}&=v_0+v_1\beta-kC. \label{eq:Z}
\end{eqnarray}

In this paper we use the parameter values introduced in~\cite{Dupont1993} as listed in Table~\ref{tab:parameters}. The only parameter that we change with respect to those of~\cite{Dupont1993} is $\beta$, since the value $\beta=0.4$ used in that paper yields oscillations while we are after an excitable behavior. 
\begin{table}[ht!]
    \centering
    \begin{tabular}{|c|c|c|c|c|}
    \hline {\bf Parameter} & {\bf Value} \\ \hline $v_0$  & $3.4\mu M min^{-1}$
    \\ \hline $v_1$  & $3.4\mu M min^{-1}$ \\ \hline $v_{M2}$ & $50\mu M min^{-1}$ \\ \hline
    $v_{M3}$   & $650\mu M min^{-1}$ \\ \hline $K_A$  & $0.9\mu M$
    \\ \hline $K_2$  & $1\mu M$ \\ \hline
    \hline $K_R$  & $2\mu M$ \\ \hline
 $k_f$  & $1 min^{-1}$ \\ \hline
 $k$  & $10 min^{-1}$  \\ \hline
 $p$  & $4$ \\ \hline
 $m$  & $2$ \\ \hline
 $n$  & $2$\\
    \hline
    \end{tabular}
    \caption{Model parameters that we use in this paper, as introduced in~\cite{Dupont1993}.}
    \label{tab:parameters}
\end{table}
As explained in what follows, in the model, increments in stimulus strength are 
represented by increments in $\beta$.  The aim of our calculations here is to determine the
dependence with $\beta$ of the probability per unit time that a spike
is elicited for values of $\beta$ for which the dynamical system is
excitable. Relating $\beta$ with [\ip] and, eventually, with the
external stimulus strength, $I_{ext}$, we expect to determine the conditions
under which this probability is an exponential function of $I_{ext}$.
Under these conditions,  the occurrence of a
spike will be a Poisson process with an average inter-spike period that
decreases exponentially with the stimulus strength.

\section{Analysis of the model}

{
\subsection{External stimuli and the parameter $\beta$.}

Eqs.~(\ref{eq:C})--(\ref{eq:Z}) model the dynamics of intracellular
\ca\ when \ca\ release through \iprs\ is involved. As already
mentioned, \iprs\ need to bind (cytosolic) \ca\ and \ip\ to become
open and, normally, \ip\ is produced upon stimulation with external
effectors such as hormones or neurotransmitters.  The model does not
include an explicit description of \ip\ production. The effect of the
external stimulus is implicitely included {\it via} the parameter
$\beta$ which represents the fraction of \iprs\ with \ip\ bound that
are ready to become open upon \ca\ binding.  Increments in stimulus
strength are thus represented by increments in $\beta$.  Dupont and Goldbeter
analyzed the bifurcation diagram of the model for the parameters in Table~\ref{tab:parameters} and $\beta\in(0,1)$~\cite{Dupont1993}. They 
determined that there is a stable fixed point for $\beta \lesssim 0.24$ and for $\beta\gtrsim 0.8$
and a stable limit cycle for $\beta$ values in between. In what follows we analyze the values of $\beta$ for which the stable fixed point is excitable. }

\subsection{Equilibrium solutions and nullclines.}

\begin{figure}[ht!]
\centering
  \includegraphics[width=0.8\linewidth]{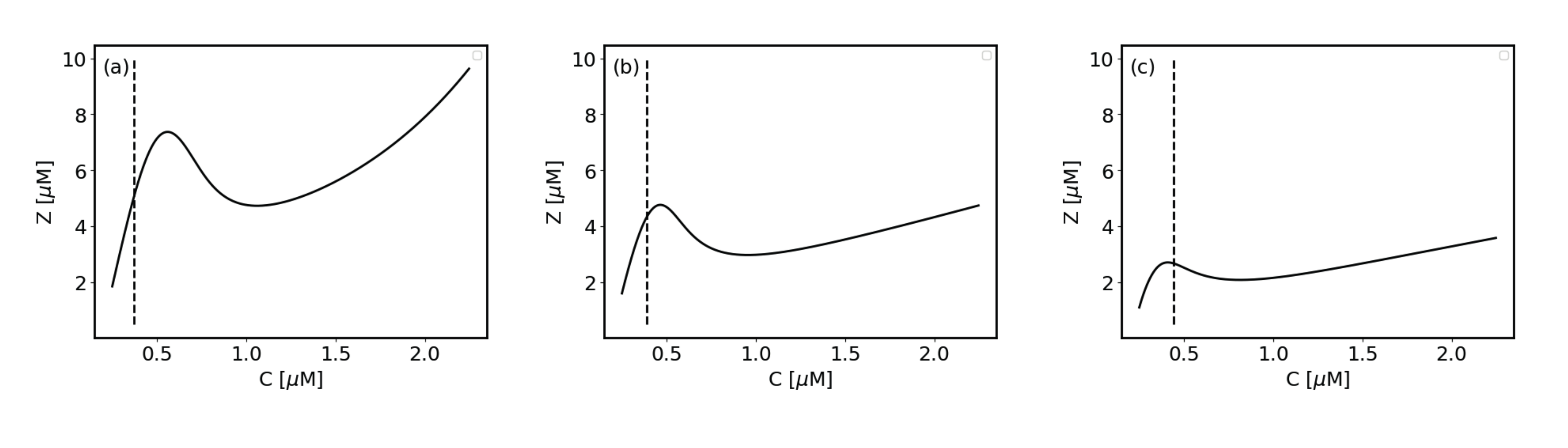}
    \caption{Nullclines of the model given by Eqs.~(\ref{eq:C})--(\ref{eq:Z}) for the parameter values of Table~\ref{tab:parameters} and $\beta =0.09$ (a), $\beta =0.15$ (b) and
      $\beta =0.3$ (c).}
    \label{fig:nullclines}
\end{figure}

\begin{figure}[ht!]
\centering
   \includegraphics[width=0.8\linewidth]{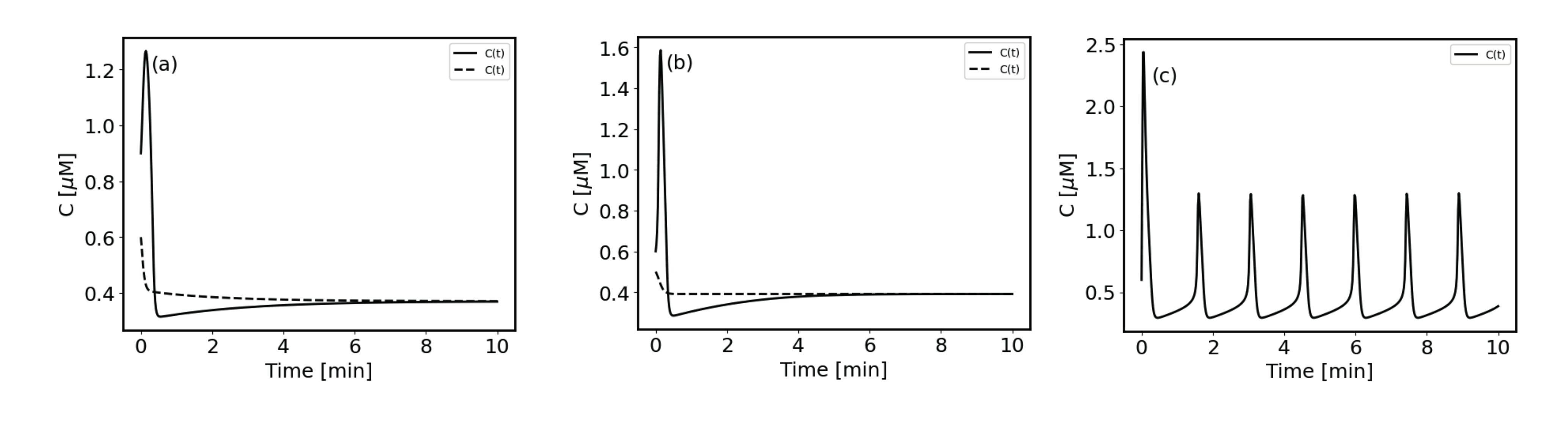}
    \caption{Time course of $C$ derived by solving Eqs.~(\ref{eq:C})--(\ref{eq:Z}) for the same parameter values as in Fig.~\ref{fig:nullclines}. In (a) and (b) we show the time evolution for two initial conditions to illustrate that the only fixed point of the system is stable but excitable. In (c) there is a stable limit cycle.  
}
    \label{fig:trajectories}
\end{figure}

We show in Fig.~\ref{fig:nullclines} the nullclines of the model for three values of $\beta$. 
We observe that for all $0\le \beta\le 1$
there is only one fixed point, $(C_{eq},Z_{eq})$, which satisfies:
\begin{equation}
C_{eq}=\frac{v_0+v_1\beta}{k}. \label{eq:Ceq}
\end{equation}
Thus, $C_{eq}$ increases with $\beta$.  {On the other hand, for all cases it is $\dot{Z}>0$ ($\dot{Z}<0$) 
to the left (right) of the nullcline depicted with dashed lines while $\dot{C}>0$ ($\dot{C}<0$) 
in the region above (below) the nullcline depicted with solid lines. This is equivalent to the nullcline structure of a prototypical example of excitability, the FitzHugh-Nagumo (FHN) model, if we identify $-Z$ and $C$ of Eqs.~(\ref{eq:C})--(\ref{eq:Z}) with the slow and fast variables of the FHN model, respectively~\cite{izhikevich_book}. Thus, as the nullclines cross to the left of the value, $C$, at which the $\dot{C}$ nullcline has its local maximum, the fixed point is excitable as illustrated in what follows.   In fact,} for the case of
Fig.~\ref{fig:nullclines}~(a), which corresponds to $\beta =0.09$, the
only fixed point is excitable: initial conditions that start on one or
the other side of the ``middle branch'' of the $\dot{C}$ nullcline
either give rise to a relatively long excursion in phase space or
decay monotonically to the fixed point. This is illustrated in
Fig.~\ref{fig:trajectories}~(a) where we have plotted the time
evolution of $C$ for $\beta =0.09$ and two such initial conditions. As
$\beta$ is increased, $C_{eq}$ gets closer to the value at which the
$\dot{C}$ nullcline has its local maximum. The equilibrium solution
continues to be excitable as long as $C_{eq}$ is bounded away from
this local maximum, {\it e.g.}, for $\beta=0.15$ for which we show the
nullclines and two trajectories in Figs.~\ref{fig:nullclines}~(b)
and~\ref{fig:trajectories}~(b), respectively.  It can be observed in
Fig.~\ref{fig:trajectories}~(b) that the excursion in phase space is
larger than in ~\ref{fig:trajectories}~(a), consistently with the
structure of the nullclines. At $\beta\approx 0.183$ the fixed point
becomes a stable focus and then loses stability at $\beta\approx
0.237$ through a Hopf bifurcation. For $0.237\lesssim \beta\lesssim
0.775$ the stable (asymptotic) solution is a limit cycle, as
illustrated in Fig.~\ref{fig:trajectories}~(c) which corresponds to
$\beta =0.3$ for which the nullclines are plotted in
Fig.~\ref{fig:nullclines}~(c). As $\beta$ gets closer from below to
$\beta =0.775$, the oscillations become more sinusoidal and of smaller
amplitude until a new Hopf bifurcation, at $\beta\approx 0.775$, makes
the fixed point stable again {for larger values of $\beta$.
For $\beta\gtrsim 0.8$, perturbations to the fixed point do not yield solutions in which $C$ increases significantly with respect to the stationary solution value. Thus, this region does not correspond to an excitable regime
in which \ca\ pulses can be elicited. Thus, for our purposes, from now on
we will restrict the analysis to $\beta\le 0.22$ to guarantee the excitability of the system.}

\begin{figure}[ht!]
\centering
   \includegraphics[width=0.3\linewidth]{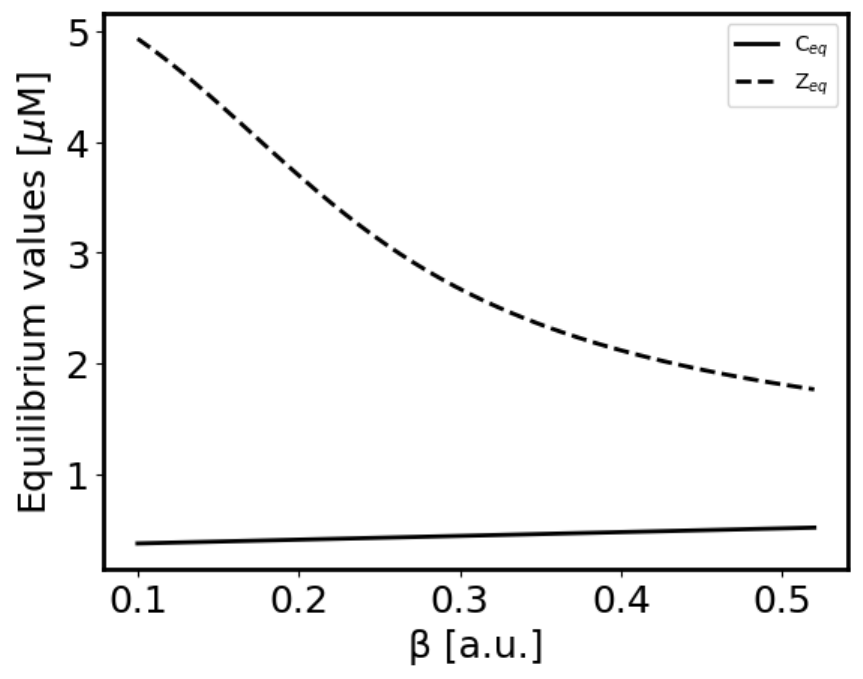}
    \caption{Equilibrium concentrations, $C_{eq}$ and $Z_{eq}$, for values of $\beta$ for which
the only equilibrium solution is excitable. The corresponding \ca\ concentration in the lumen of the ER can 
be derived from $Y_{eq} = Z_{eq}-C_{eq}$. 
}
    \label{fig:equilibria}
\end{figure}

The simple model provides reasonable values of $C_{eq}$ and $Y_{eq}$
for the range of parameters for which the only stable solution is
excitable.  Namely, in most cell types, the free cytosolic
\ca\ concentration is very small at basal levels ($\sim
50-100nM$~\cite{piegari_2014}), consistent with the values depicted in
Fig.~\ref{fig:equilibria} (which satisfy $C_{eq}\sim 0.4\mu M$
throughout the figure).  In {\it X. Laevis} oocytes, the free
\ca\ concentration in the lumen of the ER was estimated to be $\sim
200\mu M$, the ratio of the luminal to the cytosolic volume as 0.7 and
that of the fraction of free \ca\ in the cytosol and in the ER as
0.00125~\cite{Lopez_2016}. In a mechanistic interpretation of
Eqs.~(\ref{eq:C_1})--(\ref{eq:Y}), $Y$ could be identified with the
product of these two ratios ($\sim 0.009$) times the free luminal
\ca\ concentration. Thus, equilibrium values, $Y_{eq}\sim 3.5-4\mu M$,
as those that can be derived from Fig.~\ref{fig:equilibria}, are
consistent with the estimates of~\cite{Lopez_2016}.

\subsection{Slow and fast variables}

The equilibrium values, $C_{eq}$ and $Z_{eq}$, displayed in
Fig.~\ref{fig:equilibria} differ by about an order of magnitude. On
the other hand, during the (relatively short) pulses elicited in $C$ such as the one illustrated in Fig.~\ref{fig:trajectories}~(b), $C$ increases by a factor smaller than 4. As it is standard for the analysis of reactions with enzymes
in the quasi-stationary approximation~\cite{keener_sneyd_I}, we
introduce dimensionless concentrations, $\overline{C}=C/C_{eq}$ and
$\overline{Z}=Z/Z_{eq}$ and time, $\tau = t v_0/C_{max}$, to rewrite
Eqs.~(\ref{eq:C})--(\ref{eq:Z}) as:
\begin{eqnarray}
\frac{d\overline{C}}{d\tau}&=1+\frac{v_1}{v_0}\beta-\frac{v_{M2}}{v_0}\frac{\overline{C}^n}{\overline{K}_2^n+\overline{C}^n}+\beta \frac{v_{M3}}{v_0}\frac{\overline{C}^p}{\overline{K}_A^p+\overline{C}^p}\frac{(\overline{Z}/\epsilon-\overline{C})^m}{\overline{K}_R^m+(\overline{Z}/\epsilon-\overline{C})^m}\nonumber\\&\quad+\frac{k_f}{v_0}(\overline{Z}/\epsilon-\overline{C})-\frac{kC_{max}}{v_0}\overline{C},\label{eq:C_less}\\
\frac{d\overline{Z}}{d\tau}&=\epsilon\left(1+\frac{v_1}{v_0}\beta-\frac{k C_{max}}{v_0}\overline{C}\right), \label{eq:Z_less}
\end{eqnarray}
where $\epsilon\equiv C_{eq}/Z_{eq}\ll 1$. It is clear from Eqs.~(\ref{eq:C_less}-\ref{eq:Z_less}) 
 that $\overline{Z}$ (and, thus, $Z$) is the slow variable. This time scale separation is further
reinforced by the choice of parameter values of Table~\ref{tab:parameters} which are such that $v_{M3}$, that only affects the dynamics of $C$, is a much faster rate than all the others.

\section{Spike occurrence and interspike time average.}

We study the probability of spike occurrence per unit time restricting the analysis to the dynamics on the fast manifold. Namely, we fix $0.1\le\beta\le 0.22$, $Z=Z_{eq}(\beta)$ 
and work with the equation:
\begin{eqnarray}
  \dot{C}&=v_0+v_1\beta-v_{M2}\frac{C^n}{K_2^n+C^n}+\beta v_{M3}\frac{C^p}{K_A^p+C^p}\frac{(Z_{eq}(\beta)-C)^m}{K_R^m+(Z_{eq}(\beta)-C)^m}\nonumber\\
  &+k_f(Z_{eq}(\beta)-C)-kC \equiv -\frac{dV_\beta}{dC}, \label{eq:fast}
\end{eqnarray}
where we have introduced the definition of the potential, $V_\beta(C)$.
The rate of spike occurrence derived from the analysis of Eq.~(\ref{eq:fast}) will give the stochastic component of the interspike time. The total interspike time will then be the sum of this stochastic component, the spike duration and the recovery time (which is not accounted for by Eq.~(\ref{eq:fast})).

As already explained, $v_0$ and $k_f Y=k_f(Z_{eq}-C)$ in Eq.~(\ref{eq:fast})
represent an influx of \ca\ in the cytosol from the extracellular
medium and the ER, respectively, that is independent of cytosolic \ca\ and \ip. There is an additional \ip-dependent
term, $v_1\beta$, which was included to account for the observation
 that the mean cytosolic [\ca] increases with the level
of stimulation~\cite{Dupont1993}.  All these processes are affected by
noise, since they are the net result of localized events of
\ca\ entry. {Furthermore, as explained later in more detail, \ca\ spikes correspond to waves that propagate throughout the cell which
are preceded by spatially localized \ca\ release events from the ER (puffs). As discussed in  Sec.~\ref{sec:experiments}, puffs
can be accounted for as a noise term added to the dynamic
equation of the spatially averaged \ca\
concentration, $C$. In fact, in that Section we use puffs data to estimate
noise amplitude.}
The \ip\ produced as a consequence of the external stimulus can also fluctuate so that $\beta$ is subject to random fluctuations as well. 
Assuming that, for each level of external stimulation, $I_{ext}$, fluctuations in $\beta$, $\delta \beta(I_{ext})$,  satisfy $\delta \beta (I_{ext})\ll \langle \beta\rangle (I_{ext})$, we can embrace all the noisy part of these processes in a single
additive term and rewrite Eq.~(\ref{eq:fast}) as:
\begin{eqnarray}
  \dot{C}&=-\frac{dV_\beta}{dC} + \zeta, \label{eq:V}
\end{eqnarray}
with $\beta\approxeq \langle \beta\rangle(I_{ext})$. {For simplicity, we further approximate} $\zeta$ as a
Gaussian white noise term with $\langle \zeta(t)\zeta(t')\rangle= 2 D
\delta(t-t')$ where $D$ can, in principle, vary with $\beta$. {
  We discuss later to what extent it is reasonable to assume that $\delta \beta (I_{ext})\ll \langle \beta\rangle (I_{ext})$  and the implications of the $\beta$
dependence of the coefficient, $D$}.

\begin{figure}[ht!]
\centering
  \includegraphics[width=0.8\linewidth]{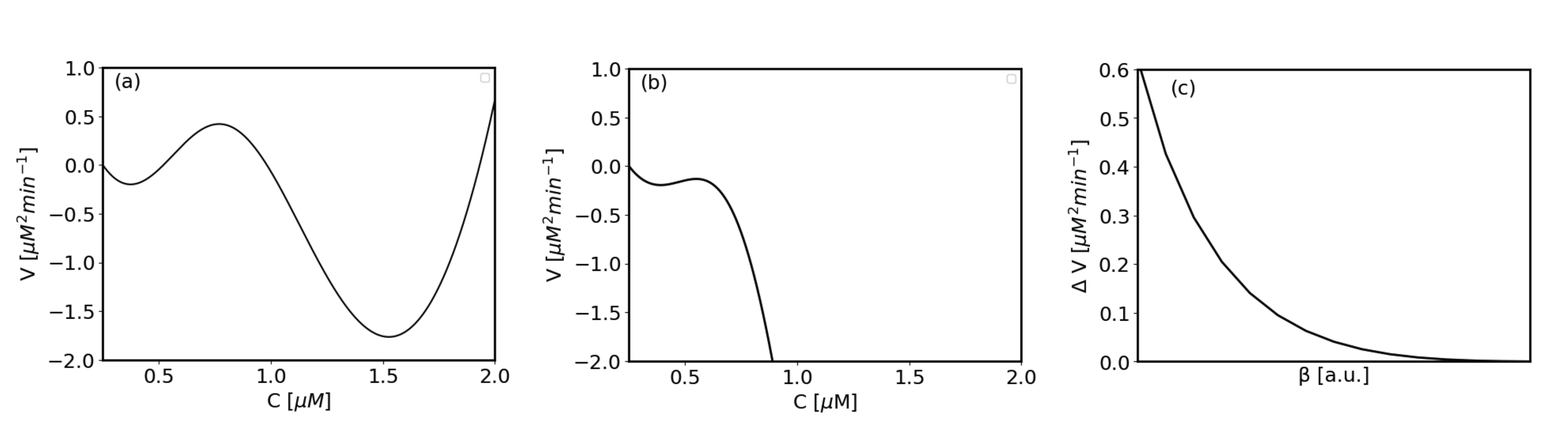}
    \caption{The potential, $V_\beta$, that determines the dynamics on the fast manifold (Eq.~(\ref{eq:V})) for $\beta=0.09$ (a) and $\beta =0.15$ (b) and the height of the barrier, $\Delta V_\beta \equiv V_\beta(C_S)-V_\beta(C_1)$, as a function of $\beta$, where $C_S$ corresponds to the local maximum of $V_\beta$ and $C_1$ to the local minimum to the left of $C_S$ (which is the value at the equilibrum solution of Eqs.~(\ref{eq:C})--(\ref{eq:Z})). }
    \label{fig:V}
\end{figure}

We show in Figs.~\ref{fig:V}~(a) and (b) a plot of $V_\beta(C)$ for the same values of $\beta$ as in Figs.~\ref{fig:nullclines}~(a) and (b). 
We observe that, in both cases, there are two local minima, $C_1$ and $C_2$, that correspond to stable fixed points of Eq.~(\ref{eq:V}) and between them a saddle point, $C_S$, that determines the separatrix between basins of attraction (on the fast manifold). One of the stable fixed points of Eq.(\ref{eq:fast}) is $C_1=C_{eq}(\beta)$, {\it i.e.}, corresponds to the stationary (excitable) solution of the full (two-dimensional) model (Eqs.~(\ref{eq:C})-(\ref{eq:Z})). The difference, $\Delta V_\beta\equiv V_\beta(C_S)-V_\beta(C_1)$, is the height of the barrier that needs to be surpassed for the system (restricted to the fast manifold) to go from $C_1$ to $C_2$. In the actual model, given by Eqs.~(\ref{eq:C})-(\ref{eq:Z}), if $\beta \lesssim 0.22$, surpassing this barrier implies that the system undergoes a long excursion in phase space, therefore, a spike is elicited.   This height decreases with $\beta$ as shown in Fig.~\ref{fig:V}~(c) a plot of $\Delta V_\beta$. 

For a bistable $V_\beta$ (as in Figs.~\ref{fig:V}~(a)-(b)) Eq.~(\ref{eq:V}) corresponds to the paradigmatic equation analyzed by Kramers~\cite{Kramers} who obtained that the time it takes for the system to leave the vicinity of the local minimum, $C_1$ (which corresponds to the excitable fixed point of Eqs.~(\ref{eq:C})--(\ref{eq:Z})) is exponentially distributed with escape rate, $r_\beta$, that, in the
low-noise limit ($D\ll \Delta V_\beta$), can be approximated by:
\begin{equation}
  r_\beta = \frac{\left( V''_\beta(C_1)\vert V''_\beta(C_S)\vert\right)^{1/2}}{2\pi} \exp (-\Delta V_\beta/D),\label{eq:rbeta}
\end{equation}
with $V''_\beta = d^2 V_\beta/dC^2$. 
Thus, the mean interspike time, $\langle T_{IS}\rangle$, would be:
\begin{equation}
  \langle T_{IS}\rangle = T_{min} + \frac{2\pi}{\left( V''_\beta(C_1)\vert V''_\beta(C_S)\vert\right)^{1/2}} \exp (\Delta V_\beta/D), \label{eq:T_IS}
  \end{equation}
with $T_{min}$ the sum of the spike duration and the recovery time.
$\langle T_{IS}\rangle$ depends exponentially on $\Delta V_\beta$ which
is a decreasing function of $\beta$ (see Fig.~\ref{fig:V}~(c)) and, thus, of
the external stimulus strength, $I_{ext}$. In the following Section we combine a variety of previously published results to derive the dependence of $\langle T_{IS}\rangle$ with $I_{ext}$ within the framework of
the very simple model considered in this paper.

\section{Interspike time average and external stimulus strength}
\label{sec:experiments}

\begin{figure}[ht!]
\centering
  \includegraphics[width=0.8\linewidth]{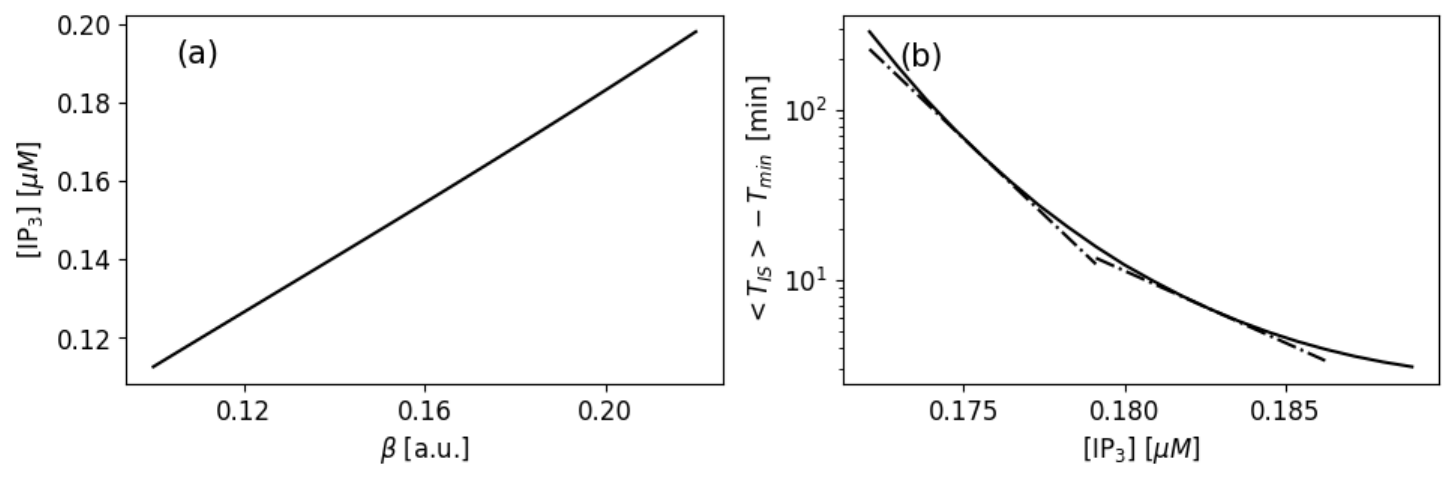}
  \caption{(a) [\ip] as a function of the \iprs\ saturation level, $\beta$, given by Eq.~(\ref{eq:beta}) with $K_{IP}=0.13\mu M$. (b) Mean stochastic part of the interspike time, $\langle T_{IS}\rangle-T_{min}$ as a function of [\ip] obtained by combining Eqs.~(\ref{eq:beta}) and (\ref{eq:T_IS}) with
    $D=0.0015\mu M^2/min$ and $K_{IP}=0.13\mu M$. Two straight lines tangent to the plotted
  curve at [\ip]$\approx 0.176\mu M$ and at [\ip]$\approx 0.183\mu M$ are also shown.  }
    \label{fig:TIS}
\end{figure}

In order to determine the dependence of $\Delta V_\beta$ with
$I_{ext}$ we need to estimate the noise level (which enters through
$D$ in Eq.~(\ref{eq:T_IS})) and the relation between $\beta$ and
$I_{ext}$. For the latter we need to relate $\beta$ to the
\ip\ concentration. We recall here that $\beta$ is the fraction of
\iprs\ with \ip\ bound that are ready to become open upon
\ca\ binding. To estimate $\beta$ we use the DeYoung-Keizer
model~\cite{dy_keizer} which considers the tetrameric structure of the
channel and assumes that three of the monomers need to have \ip\ bound
and the \ca\ inhibitory site free to eventually become open (upon
\ca\ binding to the activating site in the three monomers with
\ip\ bound). The dissociation constant of the \ip-binding unbinding
reaction depends on whether \ca\ is bound or not to the inhibitory
site. Assuming that immediately before the spike is elicited [\ca] is
approximately equal to its low basal value so that the 
\ca\ inhibitory site of the receptor is very likely free of \ca, we use the
dissociation constant that corresponds to the \ca\ free situation,
$K_{IP} = 0.13\mu M$, and relate $\beta$ and [\ip] as:
\begin{equation}
  \beta = \left(\frac{[{\rm IP}_3]}{[{\rm IP}_3]+K_{IP}}\right)^3. \label{eq:beta}
\end{equation}
We show in Fig.~\ref{fig:TIS}~(a) a plot of [\ip] $vs$ $\beta$ derived from
Eq.~(\ref{eq:beta}) with $K_{IP}=0.13\mu M$. We observe that the relation between
[\ip] and $\beta$ is approximately linear over the range of $\beta$ values for
which the dynamics of the simple model (Eqs.~(\ref{eq:C})--(\ref{eq:Z})) is excitable {with slope $d\beta/d$[IP$_3]\sim 1.4$. On the other hand, the range of [\ip] values corresponds to having
  $\sim$100 molecules/$\mu m^3$. Cell sizes vary over various orders of magnitude. The volume of the cells used in the experiments in which the exponential scaling between
  external stimulus strength and interpulse times was observed (hepatocytes and HEK293 cells~\cite{Thurley2014}) are between $10^3$ and $3\, 10^4\mu m^3$. Thus, according to our estimates, in the excitable regime,  the number of \ip\ molecules in these cells is between $10^5$ and $3\, 10^6$. Using Poisson statistics
  to estimate the ratio between the fluctuations and the mean of this number we conclude that $\delta \beta/\langle \beta\rangle$, which is equal to this
  ratio, ranges between 4\, $10^{-4}$ and 3\, $10^{-3}$.  Thus, according
  to these estimates, it is reasonable to assume that $\delta \beta \ll \langle \beta\rangle$ and that the effect of noise on the model can be approximated
by an additive term in Eq.~(\ref{eq:C}).}

To estimate the noise level {at fixed $\beta$}, we use the
experiments of~\cite{Marchant2001} in which the authors analyzed the
frequency and amplitude of the \ca\ puffs (spatially localized
\ca\ signals due to the release through one cluster of \iprs) that
occurred before a global \ca\ wave (a spike) in {\it Xenopus Laevis}
oocytes in the presence of different constant values of intracellular
[\ip]. For our calculation we assume that the \ca\ puffs are the main
random release events that enter in the noise term, $\zeta$.  In the
experiments~\cite{Marchant2001}, the \ca\ signals were observed using
the \ca\ dye, Fluo4, and confocal microscopy scanning a $65 \mu m
\times 65\mu m$ region. The fluorescence associated to each puff was
collected from a $8 \mu m \times 8\mu m$ region. For the analysis, the
authors separated the observations depending on the typical mean
interspike time, which, as expected, decreased as the amount of
released \ip\ increased, and could analyze the puffs for intermediate
(between 25 and 50$s$) and high (mostly, between $50s$ and $70$s)
values of $\langle T_{IS}\rangle$. In both situations they observed
that the amplitude of the puffs remained approximately the same. They
measured the amplitude in terms of the relative increase of the
fluorescence with respect to is basal level, $\Delta F/F_0$, obtaining
$\Delta F/F_0\sim 1$ in most cases. This implies that the local
increase in [\ca] due to a single puff was of the order of the basal
\ca\ concentration ($0.4\mu M$ in our simple model, see
Fig.~\ref{fig:equilibria}). Now, in our model
(Eqs.~(\ref{eq:C})--(\ref{eq:Z})), $C$ corresponds to the
concentration averaged over all the cell. Considering the $8 \mu m
\times 8\mu m$ region that was used to monitor puffs
in~\cite{Marchant2001}, we assume that the increase in the
cell-averaged concentration due to a single puff is $\sim (8/16)^2\,
0.4\mu M\approx 0.006\mu M$. Then, in the case of intermediate
interspike times (mean values in $25-50\, s$), puffs start to occur
approximately after 25\% of the interspike time has elapsed ({\it
  i.e.}, during 19- 38 $s$) and the number of puffs is $\sim 1000$ so
that the number of puffs per unit time is $\sim (26 - 53)
/s$. Therefore, on average, the noise term amplitude is $\zeta \sim
(26-53) 0.006\mu M/s = (0.16-0.32)\mu M/s$. For the Kramers
calculation we assumed that $\langle\zeta(t)\zeta(t')\rangle=2 D
\delta (t-t')$. Puffs are not instantaneous, so that we will
approximate $\delta (t-t')$ by the puff duration. Here we have two
options, in one case, to take into account the duration of the release
of \ca\ or of the local \ca\ elevation (estimated, respectively, as
$20$ and $150ms$ in~\cite{Bruno_2010}).  Using the latter, we then
estimate $D \sim (0.16^2-0.32^2)\mu M^2/s^2\times 0.15s /2=
(0.002-0.007)\mu M^2/s = (0.13-0.46)\mu M^2/min$.  Now, in the
experiments with large mean interspike time (most of them between
$50s$ and $70s$), the number of puffs between spikes is about 120 and
although the majority occurs after 25\% of the interspike time has
elapsed ({\it i.e.}, during 38 - 45$s$), there is a non-negligible
fraction that occurs before. Thus, the number of puffs per unit time
is $\sim (2 - 3) /s$, much smaller than for intermediate values of
$\langle T_{IS}\rangle$, while the amplitude remains approximately
equal. Repeating the calculation for this smaller puff frequency we
obtain $\zeta \sim (2-3) 0.006\mu M /s= (0.012-0.018)\mu M/s$ which
yields $D\sim (1.1- 2.4)10^{-5}\mu M^2/s = (0.0007-0.0015) \mu
M^2/min$. We then observe that $D$ increases quite noticeably
({from $\sim 0.0007\mu M^2/min$ to $\sim 0.5\mu M^2/min$}) with the
level of stimulation, {\it i.e.}, with $\beta$. {If we restrict the calculation to either the medium or the low
  stimulation level experiments, however, $D$ varies by at most a factor of
  2. It is thus reasonable to consider a fixed $D$ value in such a
  case.} Now, for the Kramers calculation that yields
Eq.~(\ref{eq:T_IS}) to hold it is necessary that $D\ll \Delta
V_\beta$. This condition is not satisfied for most of the $\beta$
values of Fig.~\ref{fig:V}~(c) if we use $D\sim(0.13-0.46)\mu
M^2/min$, the estimate derived from the experiments with intermediate
interspike times, while it is satisfied for $\beta \lesssim 0.18$ if
we use the estimate obtained for long interpike times, $D\sim
(0.0007-0.0015) \mu M^2/min$.

We show in Fig.~\ref{fig:TIS}~(b) the mean of the stochastic part of
the interspike time, $\langle T_{IS}\rangle-T_{min}$, given by
Eq.~(\ref{eq:T_IS}) as a function of [\ip] for $D=0.0015\mu M^2/min$,
where we have used Eq.~(\ref{eq:beta}) with $K_{IP}=0.13\mu M$ to
relate $\beta$ and [\ip]. The two straight lines drawn in the figure
illustrate that $\langle T_{IS}\rangle-T_{min}\approx
A\exp(-\lambda$[\ip]) with $\lambda \approx 414/\mu M$ for the range,
$0.172 \mu M\lesssim$[\ip]$\lesssim 0.179\mu M$ (for which
$15min\lesssim \langle T_{IS}\rangle-T_{min}\lesssim 290 min$) and
with $\lambda \approx 195 /\mu M$ for the range, $0.179 \mu
M\lesssim$[\ip]$\lesssim 0.186\mu M$ (for which $4min\lesssim \langle
T_{IS}\rangle-T_{min}\lesssim 15min$). We must recall that the curve
in Fig.~\ref{fig:TIS}~(b) was obtained using a fixed value of $D$ and
the analysis of the previous paragraph showed that $D$ increases
significantly with the stimulation level, {\it i.e.}, with
[\ip]. Using two different values of $D$ for the two [\ip] ranges
covered by the straight lines in Fig.~\ref{fig:TIS}~(b) does not
expand the exponential dependence over a larger region.  {
  Given the simplicity of the model considered here we can expect that
  the generic features of intracellular \ca\ oscillations be
  reproduced but not necessarily (all) the quantitative aspects
  obtained in particular cell types. The impossibility of a
  quantitative agreement across cell types is immediately apparent in
  the significant differences among the values of $T_{min}$ and
  $\langle T_{IS}\rangle$ obtained in each of the four combinations of
  cell type and effector probed and analyzed
  in~\cite{Thurley2014}. But even if the individual $T_{min}$ and $\langle T_{IS}\rangle$ values
  differ for the different experiments of~\cite{Thurley2014}, there
  are common features that the simple model is able to reproduce.  For example, if
  we take an [\ip] interval over which the curve in
  Fig.~\ref{fig:TIS}~(b) is approximately an exponential function of
  [\ip] and compute the ratio between the values,
  $\langle T_{IS}\rangle -T_{min} $, obtained at the borders of this
  interval we can see that this ratio is larger than the equivalent
  one for most of the experiments reported in~\cite{Thurley2014}. In
  particular, for the range $0.179 \mu M\lesssim$[\ip]$\lesssim
  0.186\mu M$, this ratio is $3.75$, which is about twice as large as
  the similar ratio for the experiments performed in rat primary
  hepatocytes in the presence of vasopressin ($\sim 2$) and very close
  to the one obtained for human embryonic kidney (HEK293) cells in the presence
  of carbachol ($\sim 3.97$)~\cite{Thurley2014}. This means that the
  model is able to span a range of mean interspike periods which is
  similarly large to those observed in experiments in terms of the
  characteristic times of each cell type and effector probed. Perhaps
  the range of [\ip] values over which this behavior is encountered
  with the simple model is relatively small compared with the real
  system (see discussion later). However, the ability to reproduce the
  range of mean interspike times is particularly relevant for the
  purpose that motivated the present work: unveiling the properties
  that characterize the encoding of external stimuli strengths in the
  frequency of pulsatile intermediaries of the signaling pathway.  }

\begin{figure}[ht!]
\centering
  \includegraphics[width=0.8\linewidth]{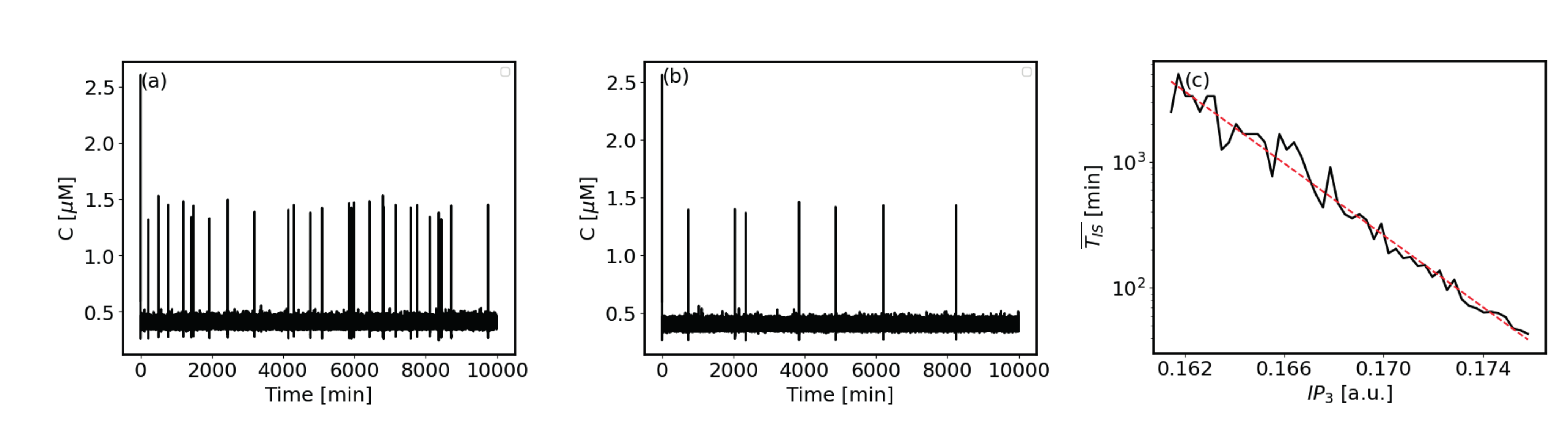}
  \caption{{Results obtained with stochastic simulations of Eqs.~(\ref{eq:C})--(\ref{eq:Z}) with the parameter values of Table~\ref{tab:parameters}, different values of $\beta$ and a 
    noise term, $\zeta$, added to Eq.~(\ref{eq:C}) as in Eq.~(\ref{eq:V}), with  $D=0.0032\mu M^2/min$. Actually the corresponding It\^{o} equations were solved using the {\it itoint} function of the {\it sdeint} package.  (a), (b) Examples of $C$ $vs$ time obtained for $\beta = 0.19$ ({\it i.e.}, [\ip] = $0.176\mu M$) and $\beta = 0.18$ ({\it i.e.}, [\ip] = $0.169\mu M$), respectively.  (c) Average  interspike time, $\overline{T_{IS}}$, derived from one 10,000 min-long simulation for each $\beta$, as a function of [\ip] (solid line). The red dashed-line corresponds to an expression of the form
$\overline{T_{IS}}=61 min \exp(-329[{\rm IP}_3]/\mu M)$. } }
    \label{fig:stoch_2D}
\end{figure}

{In spite of the above discussion, we might still ask to
  what extent the derived exponential dependence of $\langle
  T_{IS}\rangle$ on [\ip] persists if the reduction to the fast
  manifold (Eq.~(\ref{eq:V})) and the theoretical expression,
  Eq.~(\ref{eq:T_IS}), are not used. To answer this question, we
  decided to perform stochastic simulations of
  Eqs.~(\ref{eq:C})--(\ref{eq:Z}) with a noise term, $\zeta$, added to
  Eq.~(\ref{eq:C}) as previously done with Eq.~(\ref{eq:V}). The
  rationale for adding a noise term solely to Eq.~(\ref{eq:C}) is
  based on the assumption that \ca\ puffs are the main source of
  fluctuations, an assumption we have already used to estimate the
  noise amplitude from experiments. \ca\ puffs would add to the term
  proportional to $v_{M3}$ in Eq.~(\ref{eq:C}) which is not present in
  Eq.~(\ref{eq:Z}).  On the other hand, other random events of
  \ca\ entry from the extracellular medium (that would add to the
  terms proportional to $v_0$ or $v_1$ which are present in
  Eq.~(\ref{eq:Z})) would be of smaller size and their effect would be
  negligible on $Z$ (which is related to luminal [\ca] and typically
  much larger than $C$, see, {\it e.g.}, the discussion on the effect
  of \ca\ puffs on luminal \ca\ in~\cite{Lopez_2016}). We show in
  Fig.~\ref{fig:stoch_2D}~(a),(b) two examples of $C$ $vs$ time
  obtained with the stochastic simulations performed using the {\it
    itoint} function of the {\it sdeint} package with $D=0.0032\mu
  M^2/min$ and $\beta = 0.19$ ([\ip] = 0.176$\mu M$) and $\beta =0.18$
  ([\ip] = 0.169$\mu M$), respectively.  We observe in both cases a
  sequence of randomly separated \ca\ spikes (of $\sim 0.4 min$
  duration) where the mean interspike time is larger for the case with
  smaller $\beta$. We ran simulations of this type for a total
  simulation time of 10,000$min$ and various values of $\beta$ from
  which we computed the average interspike time, $\overline{T_{IS}}$,
  as a function of $\beta$.  We show in Fig.~\ref{fig:stoch_2D}~(c)
  the plot of $\overline{T_{IS}}$ as a function of [\ip] where the
  exponential dependence is apparent, as reflected by the red dashed
  straight line which corresponds to an expression of the form
  $\overline{T_{IS}}=61 min \exp(-329[{\rm IP}_3]/\mu M)$. Given the
  typical values of $\overline{T_{IS}}$ and spike duration ($\sim 0.4
  min$), it is reasonable to assume that
  $\overline{T_{IS}}-T{min}\approx \overline{T_{IS}}$, so that
  Figs.~\ref{fig:stoch_2D}~(c) and \ref{fig:TIS}~(b) are readily
  comparable, although it must be noticed that the former was obtained
  for a larger noise level because we were not restricted by the
  constraints of the theory.  We observe that also in the case of
  Fig.~\ref{fig:stoch_2D}~(c) a small change in [\ip] results in a
  relatively large range of $\overline{T_{IS}}$ values as obtained in
  Fig.~\ref{fig:TIS}~(b). The similarity is further reflected in the prefactor
  inside the exponential (329/$\mu M$ in Fig.~\ref{fig:stoch_2D}~(c)
  $vs$ $414/\mu M$ and $195 /\mu M$ in Fig.~\ref{fig:TIS}~(b)). }

In the experiments, the exponential dependence was found between the
mean of the stochastic part of the interspike time and the external
ligand concentration (vasopressin or carbachol in the two examples
described in the previous paragraph).  We should then ask whether the
exponential dependence that we found between $\langle
T_{IS}\rangle-T_{min}$ and [\ip] could also hold for the external
stimulus strength, $I_{ext}$. {As mentioned in the Introduction,
  \ip\ is produced as a result of the binding of external effectors to
  specific plasma membrane receptors.} The external stimulus strength
in our description is thus proportional to the external effector
concentration. The rate of \ip\ production is proportional to the mean
number of effector-bound receptors which, in equilibrium, is typically
a Hill function of the effector concentration. \ip, in turn, is
degraded at a rate $\sim 0.11-0.14 s^{-1}$~\cite{wang_1995}. Thus, we
can assume that, {upon constant stimulation}, [\ip] and $I_{ext}$ are related by a nonlinear
function of the form:
\begin{equation}
[{\rm IP}_3] = [{\rm IP}]_0\frac{I_{ext}^h}{I_{ext}^h+ K_D^h}, \label{eq:ip}
\end{equation}
where $h=1$ for single site receptor binding of the extracellular
effector as considered in~\cite{wang_1995}.  Now, in our simple
setting, the range of [\ip] over which the relation between $\langle
T_{IS}\rangle-T_{min}$ and [\ip] is approximately exponential ($[0.172
  \mu M, 0.179\mu M]$ or $[0.179 \mu M, 0.186\mu M]$ in
Fig.~\ref{fig:TIS}~(b)) (and for which the range of mean interspike
pulses is relatively wide compared with the experiments) entails at
most a 4\% variation in [\ip].  Thus, a linearization of
Eq.~(\ref{eq:ip}) would be relatively good in each of these ranges
guaranteeing the exponential dependence between $\langle
T_{IS}\rangle-T_{min}$ and $I_{ext}$ as well. We could still wonder
whether the function Eq.~(\ref{eq:ip}) is not saturated ($I_{ext}\gg
K_D$). It is unlikely that the saturation of the \ip\ production be
achieved for a situation with less than 20\% of \ipr\ occupation
(represented by $\beta$ in our model).  On the other hand, [\ip] was
estimated in {\it X. Laevis} oocytes to vary between 40$nM$ at basal
conditions and $\sim 1.8\mu M$ upon strong
stimulation~\cite{LUZZI1998}.  It is then reasonable to assume a
linear relation between [\ip] and $I_{ext}$ ([\ip]$\approx [{\rm
    IP}]_0 I_{ext}/K_D$ from Eq.~(\ref{eq:ip}) with $h=1$) for
situations, as those analyzed in this paper, in which [\ip]$<0.2\mu
M$.  {We can still ask how realistic it is that a relative
  change in the mean interspike time as the one observed in the
  experiments be caused by a difference of a few nM in [\ip] as
  implied by Figs.~\ref{fig:TIS}~(b) and \ref{fig:stoch_2D}~(c).  To answer this, we recall that
  the experiments performed in N1E-115 neuroblastoma cells under 30$s$
  stimulation with carbachol (CCh)~\cite{wang_1995} gave estimates of
  [\ip]$\sim 10nM$ at basal levels and [\ip]$\sim 35nM$ for
  [CCh]=1$mM$. According to the analyses of~\cite{Thurley2014}, the
  mean interspike time obtained for HEK293 cells stimulated with
  constant CCh$\in[30,200]\mu M$ depended on [CCh] as $\langle
  T_{IS}\rangle = T_{min} + B \exp(-\gamma [{\rm CCh}])$ with $\gamma
  = 7.84/mM$. Assuming a linear relation between [CCh] and [\ip], we
  can estimate [\ip] $=10nM + 25nM [{\rm CCh}]/mM$ from the
  experiments of~\cite{wang_1995}. Assuming that a similar relation
  holds for the experiments performed in HEK293 cells stimulated with
  CCh~\cite{Thurley2014} and using this expression to replace [CCh] in
  terms of [\ip] in the function that relates $\langle T_{IS}\rangle$
  and [CCh] we obtain a $\sim 4.2nM$ difference between the extreme
  values of [\ip] probed in these experiments and $\langle
  T_{IS}\rangle = T_{min} + A \exp(-\lambda [{\rm IP}_3])$ with
  $\lambda = \gamma mM/(25nM) = 313.6/\mu M$, remarkably similar to
  the values obtained with our very simple model ($ 414/\mu M$ and $ 195 /\mu M$,
  for the two straight lines in
  Fig.~\ref{fig:TIS}~(b) and $329/\mu M$ for the one in Fig.~\ref{fig:stoch_2D}~(c)).  Based on this discussion we conclude that we
  have found a very plausible} mechanistic explanation for the
exponential dependence of $\langle T_{IS}\rangle - T_{min}$ with
$I_{ext}$, as observed experimentally in~\cite{Thurley2014}.

\section{Summary and conclusions}

Cells use different strategies to decode the information they receive
from the
environment~\cite{tostevin_2012,micali_pcbi_2015,givre_sci_rep_2023}. Very
often, stationary external stimuli induce oscillations in some
intermediaries of the signaling pathway with increasing frequencies as
the stimulus strength grows~\cite{Cai2008, Carbo15022017}. This {\it
  frequency encoding} strategy is very common when one of the
intermediaries is intracellular \ca, an ubiquitous signaling messenger
that is involved in many different physiological
processes~\cite{Dupont2011,Bootman2012}. {Although
  spatial non-uniformities play an important role on the dynamics of intracellular \ca, many observations are reported in terms of spatial averages of the \ca\ distribution. In this paper we focused on this type of observations, more specifically, on the} sequences of \ca\ pulses that have
been observed in a variety of cell types in which \ca\ is released
from the Endoplasmic Reticulum (ER) through IP$_3$ receptors (\iprs)
upon constant stimulation~\cite{Skupin2008}. The interpulse times have
been found to have a relatively large stochastic component with an
average that scaled exponentially with the extracellular stimulus
strength~\cite{Thurley2014}. Assuming that this behavior could be
related to the excitability of the intracellular \ca\ dynamics, in
this paper we used a simple model {in terms of ODEs} with noise to derive theoretically {and numerically} this
exponential scaling.

 In this work we used the simple model of \ipr-mediated intracellular
 \ca\ dynamics of~\cite{Dupont1993} for values of the fraction of \ip-bound \iprs\ ready
 to become open upon \ca\ binding, $\beta$, such that the system had
 only one stable and excitable stationary solution.  Relying on the
 time scale separation of the processes, we restricted the dynamics to
 the fast manifold and analyzed the probability per unit time that
 noisy perturbations could elicit a long excursion in phase space
 ({\it i.e.} a spike). Assuming that the noisy perturbations were due
 to random localized events of \ca\ entry into the cytosol which could
 be embraced in a single additive term, we used Kramers law to derive
 the said probability and the mean of the stochastic part of the
 interspike time, $\langle T_{stoch}\rangle\equiv \langle
 T_{IS}\rangle-T_{min}$. In this way we obtained an exponential
 dependence between $\langle T_{stoch}\rangle$ and the height of the
 (potential) barrier that the system had to cross to produce a spike,
 $\Delta V_\beta$, which, in turn, was a decreasing function of $\beta$
 (Fig.~\ref{fig:V}~(c))).  Using the DeYoung-Keizer
 \ipr\ model~\cite{dy_keizer} to relate $\beta$ and [\ip]
 (Eq.~(\ref{eq:beta})) and estimating the noise level from experimental
 observations~\cite{Marchant2001}, we determined the relation between
 $\langle T_{stoch}\rangle$ and [\ip] (Fig.~\ref{fig:TIS}~(b)).  In
 particular, we found that this relation could be approximated by an
 exponential over different [\ip] ranges (see the two straight lines
 in Fig.~\ref{fig:TIS}~(b)) that produced relatively wide $\langle
 T_{stoch}\rangle$ intervals (of the order of or larger than those
 observed experimentally~\cite{Thurley2014}).  Based on previous
studies on the amount of \ip\ produced upon external stimulation in 
different cell types~\cite{wang_1995,LUZZI1998} we could argue that [\ip] was linearly proportional
to the external stimulus stregnth, $I_{ext}$, finding in this way a  mechanistic explanation for the
exponential dependence of $\langle T_{IS}\rangle - T_{min}$ with
$I_{ext}$, observed in experiments~\cite{Thurley2014}.

{The mechanistic explanation was initially derived
  restricting the dynamics of the model system to its fast manifold
  (Eq.~(\ref{eq:fast})) and computing the mean interpulse time using
  Kramers theory (Eq.~(\ref{eq:T_IS})), which is valid in the low
  noise limit. This strategy has been used to analyze the dynamics of
  excitable systems when noise is added to the fast variable equation
  in the context of Self-Induced Stochastic Resonance
  (SISR)~\cite{SISR,SISRvsCR}. Differently from the present paper
  where the interest lies on a regime where the sequence of spikes has
  a large stochastic component, these previous studies focused on the
  emergence of (quasi-regular) limit cycles due to noise. To the best
  of our knowledge, previous works have not looked at the aspect which
  motivated the present study: the exponential dependence between
  interspike time and stimulus strength, particularly, within the
  framework of cell signaling. We focused on this aspect because it
  endows cells with information transmission capabilities that are
  qualitatively different from other commonly used strategies of
  stimuli encoding~\cite{Givre2023_2}.  We also studied to what extent
  the exponential dependence of $\langle T_{IS}\rangle$ on $I_{ext}$
  or equivalently, on $\beta$ or [\ip], persisted if the reduction to
  the fast manifold and Kramers theory were not used. To this end, we
  performed stochastic simulations of Eqs.~(\ref{eq:C})--(\ref{eq:Z})
  with a noise term, $\zeta$, added to Eq.~(\ref{eq:C}). For the simulations we
  used a slightly larger noise amplitude since we were not restricted
  by the conditions of the theoretical derivation and obtained an
  exponential dependence that seemed to hold for larger [\ip] ranges
  than those obtained with the analytic derivation
  (Fig.~\ref{fig:stoch_2D}). The results obtained with the same noise
  level as in Fig.~\ref{fig:TIS}~(b) were quantitatively similar to
  those of Fig.~\ref{fig:TIS}~(b) (data not shown). All these figures
  illustrate that the mechanism which underlies the occurrence of
  spikes in our model is such that a relatively large range of mean
  interspike times is spanned. As noted in~\cite{SchusterStefanMarhl},
  this is particularly advantageous for frequency encoding, but while
  in~\cite{SchusterStefanMarhl} it was argued that this situation could be
  achieved as a bifurcation was approached at which the period
  diverged, in our case this behavior derives from the escape rate
  from an excitable fixed point due to noise. }

{In this paper we used a model of the intracellular
  \ca\ dynamics in terms of ODEs with noise in which the effect of the spatial
  inhomogeneities that are inherent to these signals were implicitely
  included in the noise term. Namely, as discussed in the previous
  Section, we used experimental results on the spatial inhomogeneities
  underlying \ca\ pulses to estimate the noise weight. There is always
  the possibility of using more realistic spatio-temporal models. In
  this regard, there are multiple options: intracellular \ca\ signals
  involve processes that occur over many different length and time
  scales and models can describe each of these scales with more or
  less
  detail~\cite{fdf_pnas,Calabrese_2010,Ventura_2006,Solovey_2008,Solovey_2010,Lopez2012,Voorsluijs_2019}.
 The choice of model depends on the type of question that the
  research seeks to answer and more detailed ones give more thorough
  mechanistic descriptions of the underlying processes~\cite{Thurley_2011,FRIEDHOFF2023}. The
  simple model used here, however, provided an easily interpretable
  explanation of the exponential dependence between the stimulus
  strength and the mean interpulse time which can readily be applied to other
  pulsatile signaling agents.  

A
similar approach to the one described here can be applied in many other situations} given the widespread
presence of excitable dynamics in biology.  Candidates to be described
with this approach include Transcription Factors (TFs) that display
bursts of nuclear translocation, for which there are several signs of
noise induced processes. Examples of this behavior occur in the
response to stress in yeast mediated by the TF, Msn2~\cite{msn22}, the
differentiation in bacteria that could be described correctly with a
noise-triggered excitable system~\cite{functional_role_noise} or the
(mammalian) tumor supressor, p53~\cite{p53_excitable}, which dynamics
requires an underlying excitable network
structure~\cite{p53_excitable_network_falcke}. It is relevant here to
determine whether the pulses of nuclear localization or of activation
of the intermediary of the transduction from the external stimulus to
the response scale exponentially with the stimulus strength. We have
found that this exponential dependence can endow frequency encoding
with the ability to discriminate stimulus strengths equally well over a relatively wide range
of intensities, making it qualitatively different from amplitude
encoding~\cite{Givre2023_2}.  Having a theoretical framework to
account for this dependence gives support to the assumption that this qualitative difference occurs in many different settings. This idea is supported by the fact that the mean
frequency of nuclear localization bursts of the TF, Crz1, in yeast can
be shown to increase exponentially with extracellular
\ca~\cite{Cai2008} while the frequency of other TFs are convex
increasing functions of the stimulus
strength~\cite{DALAL20142189,Hao2012} which, upon further scrutiny,
might end up being exponential functions.

%Because the characteristic times (at $T\sim 37º$) are usually characterized by $\frac{h}{kT}e^{\frac{\Delta G}{R T}}\sim 16ps \ e^{\frac{\Delta G}{0.62}}$, then 10 min gives around $\Delta G \sim 21 J/mol$. Because this is for $\delta \beta$, it would imply $8.4 kJ/mol$ to fill the $IPT_3$ from minimum to maximum.

\subsection{Acknowledgments}
This research has been supported by
UBA (UBACYT 20020170100482BA and 20020220300192BA) and ANPCyT (PICT 2018-02026 and PICT-2021-III-A-00091). SPD
thanks the organizers of StatPhys28 for their support and cooperation. 

\subsubsection{References.\label{bibby}}

\bibliographystyle{unsrt}
\bibliography{bibliography}

%You can produce your bibliography in the standard \LaTeX\ way using the \verb"\bibitem" command. Alternatively
%you can use BibTeX: our preferred  \verb".bst" styles are: 

%\begin{itemize}
%\item For the numerical (Vancouver) reference style we recommend that authors use 
% \verb"unsrt.bst"; this does not quite follow the style of published articles in our
% journals but this is not a problem.  Alternatively \verb"iopart-num.bst" created by Mark A Caprio
% produces a reference style that closely matches that in published articles.  The file is available from
%\verb"http://ctan.org/tex-archive/biblio/bibtex/contrib/iopart-num/" .
%\item For alphabetical (Harvard) style references we recommend that authors use the \verb"harvard.sty"
%in conjunction with the \verb"jphysicsB.bst" BibTeX style file.  These, and accompanying documentation, can be downloaded
%from \penalty-10000 \verb"http://www.ctan.org/tex-archive/macros/latex/contrib/harvard/".
%Note that the \verb"jphysicsB.bst" bibliography style does not include article titles
%in references to journal articles.
%To include the titles of journal articles you can use the style \verb"dcu.bst" which is included
%in the \verb"harvard.sty" package.  The output differs a little from the final journal reference
%style, but all of the necessary information is present and the reference list will be formatted
%into journal house style as part of the production process if your article is accepted for public%ation.
%\end{itemize}

%\noindent Please make sure that you include your \verb".bib" bibliographic database file(s) and any 
%\verb".bst" style file(s) you have used.

\end{document}